\documentclass[preprint,amsmath,amssymb]{revtex4}
\usepackage{graphicx}
\usepackage{dcolumn}
\usepackage{bm}

\begin{document}
\title{Diffusion constant of slowly rotating black three-brane  }
\author{  Z.~Amoozad $^{a}$ \footnote{Electronic
address:~z.amoozad@stu.umz.ac.ir }
and  J. ~Sadeghi $^a$ \footnote{Electronic address: ~pouriya@ipm.ir , corresponding author } }
\address{$^{a}$ Sciences Faculty, Department of Physics, University of Mazandaran, Babolsar, Iran \\
P.O.Box 47416-95447.}

\begin{abstract}
In this paper, we take the slowly rotating black three-brane background and perturb it by introducing a vector gauge field. We find the components of the gauge field through Maxwell equations and Bianchi identities. Using currents and some ansatz we find Fick's first law at long wavelength regime.
An interesting result for this non-trivial supergravity background is that the diffusion constant on the stretched horizon which emerges from Fick's first law is a complex constant.
The pure imaginary part of the diffusion constant appears because the black three-brane has angular momentum. By taking the static limit of the corresponding black brane the well known diffusion constant will be recovered.
On the other hand, from the point of view of the Fick's second law, we have the dispersion relation $\omega=-iDq^{2}$ and we found a damping of hydrodynamical flow in the holographically dual theory. Existence of imaginary term in the diffusion constant introduces an oscillating propagation of the gauge field in the dual field theory. 
\end{abstract}
\maketitle

\section{Introduction}
\label{intro}
In ten dimensional type IIB supergravity a black three-brane is a cluster of coincident $D3$-branes ~\cite{a}.
Black three-branes may have angular momentum in a plane perpendicular to the brane and the geometry of this object has extensively been discussed in ~\cite{b,c,d}.
Generally a three-brane in ten dimensions has a rank three transverse rotation group $SO(6)$, so it has three independent (commuting) angular momentums.

An interesting aspect of $D3$-brane is that worldvolume of $N$ coincident $D3$-branes, being in ten dimensional spacetime and at low energies, can be
 described by super-Yang-Mills theory with gauge group $U(N)$. This point shows that $D3$-branes and black three-branes might be used in the context of dual theories.
 The dual theory which connects physics in a region of space with gravity to a region without gravity; is known as holographic principle.

In the context of the holographic principle ~\cite{e,f,g}, the region without gravity which lives on smaller dimensions, can describe in particular cases some kinds of field theories. As we know, the
fluid/gravity duality connects the long wavelength field theories to a gravity dual containing a black brane with nonzero Hawking temperature ~\cite{h,i}.
 It has been proved that theories living on the non extremal $D3$, $M2$ and $M5$ branes have hydrodynamical behavior at long wavelength description and their hydrodynamic modes are coincide
  with kinetic coefficients in the dual theories extracted from AdS/CFT ~\cite{j,k,l,m,n}.

On the other hand, the membrane paradigm, which describes the hydrodynamic-like properties of the event horizon, would be a better physical model for black branes in comparison with black holes because the horizon of black holes lacks translational invariants ~\cite{rr,ss}. Also considering fluctuations around static black brane solutions, the diffusion relations and
shear flow in the holographically dual theory have been extracted in ~\cite{o,gg}.

Some properties of the rotating black three-brane  such as thermodynamic properties and stability has been studied in the literature~\cite{cc,ee,ff}. Also the relation between rotating black three-branes and $QCD$ has been calculated in ~\cite{dd}. The points mentioned above motivated us to investigate dispersion relation of dual field theory of rotating black three-brane spacetime which the hydrodynamical modes are encoded in. It helps us to understand more about fluid/gravity duality for the case of rotating branes. So we take slowly rotating black three-brane
in the gravity side and introduce a small fluctuation, using a vector gauge field from the field theory side, and find Fick's first law.
By extracting Fick's first law, an explicit expression for the diffusion constant will be obtained. Using the Fick's second law, i.e. $\omega=-iDq^{2}$, we find a traditional damping of the hydrodynamical modes (because of the real part of the diffusion constant) plus an oscillating   propagation of the gauge field, which emerges from the imaginary part of the obtained diffusion constant, in long wave length regime of the corresponding field theory.

The organization of the paper is as follows. In section II, we introduce the reduced spacetime of
rotating black three-brane. By using the gauge field, we perturb this background and then apply Maxwell equations and Bianchi identities on the stretched horizon. In that case, in order to determine all appropriate components of the gauge field
we suppose some ansatz. In section III, we will obtain Fick's first law by relating different components of Maxwell equations. Then, we extract an explicit expression for the diffusion constant of the dual field theory. Section IV is devoted to the Fick's second
law and the dispersion relation which relates the quasi normal modes of black branes to the diffusion constant. In the final section we have conclusions and notes.

\section{ Rotating black three-brane and corresponding perturbation }
\label{sec:1}
The general rotating brane solution in string theory is presented in \cite{dd,ff}. So thermodynamic properties and stability conditions of that solutions can be found easily.
One of the hydrodynamic modes is the diffusion constant and in order to find it we consider the reduced metric of the rotating three-brane with angular momentum (here we have just one
angular momentum),
\begin{eqnarray}
\label{eq:1}
ds_{IIB}^{2}&=&f^{-\frac{1}{2}}\left\{-hdt^{2}+dx_{1}^{2}+dx_{2}^{2}+dx_{3}^{2}\right\} \nonumber \\
            &+&f^{\frac{1}{2}} \left\{\frac{dr^{2}}{\widetilde{h}}+ r^{2}(\Delta d\theta^{2}+\widetilde{\Delta}sin^{2}\theta d\Phi^{2}+cos^{2}\theta d\Omega_{3}^{2})
 -\frac{2lr_{0}^{4}cosh\alpha}{r^{4}\Delta f}sin^{2}\theta dt d\Phi \right\},
\end{eqnarray}
in which,
\begin{eqnarray}
\label{eq:2}
f&=&1+\frac{r_{0}^{4}sinh^{2}\alpha}{r^{4}\Delta},   \nonumber \\
 \Delta&=&1+\frac{l^{2}cos^{2}\theta}{r^{2}},\;\;\;\;\;\;\;\;\;\;\;\;\;\;\;\ \widetilde{\Delta}=1+\frac{l^{2}}{r^{2}}+\frac{r_{0}^{4}l^{2}sin^{2}\theta}{r^{6}\Delta f}, \nonumber \\
   h&=&1-\frac{r_{0}^{4}}{r^{4}\Delta}, \;\;\;\;\;\;\;\;\;\;\;\;\;\;\;\;\;\;\     \widetilde{h}=\frac{1}{\Delta}(1+\frac{l^{2}}{r^{2}}-\frac{r_{0}^{4}}{r^{4}}),
\end{eqnarray}
and we have assumed ~\cite{dd},
\begin{equation}
\label{eq:3}
d\Omega_{3}^{2}=d\Psi_{1}^{2}+sin^{2}\Psi_{1}d\Psi_{2}^{2}+cos^{2}\Psi_{1}d\Psi_{3}^{2}.
\end{equation}
Similar to spherically symmetric cases the dilaton is constant and $cosh\alpha$ relates directly to the three-brane charge. As we know, the rotating black three-brane and $N$ coincident $D3$-branes with a density of $R$-charge on the world-volume have the same quantum numbers. Thus we have,
\begin{equation}
\label{eq:4}
r_{0}^{4}sinh\alpha cosh\alpha=R^4\equiv4\pi g_{s}N{\alpha^\prime}^{2},
\end{equation}
where $g_{s}$ and ${\alpha^\prime}$ are the string theory parameters. The event horizon for the metric (1) is,
\begin{equation}
\label{eq:5}
r_{e}^{2}=\frac{1}{2}\left\{-l^{2}+\sqrt{l^{4}+4r_{0}^{4}} \right\},
\end{equation}
where the Hawking temperature at $\theta=\frac{\pi}{2}, \Psi_{1}=0$ is given by,
\begin{equation}
\label{eq:6}
T\mid_{r=r_{e}}=\frac{r_{e}}{2\pi r_{0}^{4}cosh\alpha}(2r_{e}^{2}+l^{2}).
\end{equation}
In the slowly rotating limit (small angular momentum) we take $\frac{l^{2}}{r_{0}^{2}}\ll1$  and higher order terms can be neglected.
So $r_{e}^{2}=r_{0}^{2}(1-\frac{l^{2}}{2r_{0}^{2}})+O(\frac{L^{4}}{r_{0}^{4}})$, and the temperature is $T\mid_{r=r_{e}}=\frac{1}{\pi r_{0}cosh\alpha}(1-\frac{l^{2}}{4r_{0}^{2}})$. It is obvious that (4) connects $r_{0}$ and $cosh\alpha$ to the $D3$-brane charge $N$. So, in the low energy limit, i.e. $N\rightarrow\infty$, we have $cosh\alpha=(\frac{4\pi g_{s}{\alpha^\prime}^{2}}{r_{0}^{4}}N)^{\frac{1}{2}}=\frac{R^{2}}{r_{0}^{2}}$ and the Howking temperature will be,
\begin{equation}
\label{eq:7}
T_{H}|_{N\rightarrow\infty}=\frac{r_{0}}{\pi(4\pi g_{s}{\alpha^\prime}^{2}N)^{\frac{1}{2}}}(1-\frac{l^2}{4 r_{0}^2})=\frac{r_{0}}{\pi R^{2}}(1-\frac{l^2}{4 r_{0}^2}).
\end{equation}
This relation expresses that in the large 'tHooft coupling $(\lambda=g_{YM}^{2}N\gg1)$, Howking temperature of rotating three-brane in the static limit gives the $AdS5$ result  exactly. %Truly we expect that, because extremal three-brane in the near horizon is the low energy limit and its metric can transform to $AdS5\times S^5$ in the Poincare coordinate by tiny calculations.

The stretched horizon is a suitable region to calculate the diffusion constant on. In our case, the stretched horizon is a flat spacelike hypersurface located at $r=r_{h}$ such that,
\begin{equation}
\label{eq:8}
r_{h}>r_{e},\;\;\;\;\;\;\;\;\;\ r_{h}-r_{e}\ll r_{e}.
\end{equation}
We find in the next sections a more restricted constraint than (8).
The unit normal vector on the corresponding hypersurface is a spacelike vector. Using $\Phi=r=const.$ we have $n_{r}=\sqrt{g_{rr}}$.
Also the inverse components of the metric $ds^{2}_{IIB}$ are,
\begin{eqnarray}
\label{eq:9}
g^{tt}&=&\frac{-g_{\Phi\Phi}}{g_{t\phi}^{2}-g_{tt}g_{\Phi\Phi}},\;\;\;\;\;\
 g^{\Phi\Phi}=\frac{-g_{tt}}{g_{t\phi}^{2}-g_{tt}g_{\Phi\Phi}},\;\;\;\;\;\
 g^{t\Phi}=\frac{g_{t\Phi}}{g_{t\phi}^{2}-g_{tt}g_{\Phi\Phi}}, \nonumber\\
g^{x_{i}x_{i}}&=&(g_{x_{i}x_{i}})^{-1}, \;\;\;\;\;\;\;\;\;\;\;\;g^{rr}=(g_{rr})^{-1}, \;\;\;\;\;\;\;\;\;\;\ g^{\theta\theta}=(g_{\theta\theta})^{-1}, \;\;\;\;\;\;\;\;\;\;g^{\Psi_{i}\Psi_{i}}=(g_{\Psi_{i}\Psi_{i}})^{-1}.
\end{eqnarray}

The hydrodynamic theory in the dual field theory can be produced by currents and tensors, so it is important to investigate the vector and tensor perturbations.
The dynamics of vector perturbation can be found by Maxwell action,
\begin{equation}
\label{eq:10}
S_{gauge}\sim\int dx^{p+2}\sqrt{-g}(\frac{1}{g_{eff}^{2}}F^{\mu\nu}F_{\mu\nu}).
\end{equation}
Taking an external gauge field $A_{\mu}$ as a perturbation, one can study currents on the stretched horizon $r=r_{h}$. The corresponding equation of current is,
\begin{equation}
\label{eq:11}
J^{\mu}=n_{\nu}F^{\mu\nu}\mid_{r_{h}},
\end{equation}
after expansion we have,
\begin{eqnarray}
\label{eq:12}
J^{t}&=&\frac{g^{tt}}{\sqrt{g_{rr}}}F_{tr},\;\;\;\;\;\;\;\;\;\;\;\;\;\;\;\;J^{x_{i}}=\frac{1}{g_{x_{i}x_{i}}\sqrt{g_{rr}}}F_{x_{i}r},\;\;\;\;\;\;\;\;\;\;\;\;\;\;\;\;J^{\theta}=\frac{1}{g_{\theta\theta}\sqrt{g_{rr}}}F_{\theta r},    \nonumber\\
J^{\Phi}&=&\frac{g^{\Phi\Phi}}{\sqrt{g_{rr}}}F_{\Phi r},\;\;\;\;\;\;\;\;\;\;\;\;J^{\Psi_{i}}=\frac{1}{g_{\Psi_{i}\Psi_{i}}\sqrt{g_{rr}}}F_{\Psi_{i}r}.
\end{eqnarray}
The conservation equation of current is $\partial_{\mu}J^{\mu}=0$, this leads us to the following equation,
\begin{equation}
\label{eq:13}
\frac{g^{tt}}{\sqrt{g_{rr}}}\partial_{t}F_{tr}+\frac{1}{g_{x_{i}x_{i}}\sqrt{g_{rr}}}\partial_{x_{i}}F_{x_{i}r}+\partial_{\theta}(\frac{F_{\theta r}}{g_{\theta\theta}\sqrt{g_{rr}}})=0.
\end{equation}
Because of the antisymmetry properties of $F^{\mu\nu}$ one can conclude $n_{\mu} J^{\mu}=0$, which shows that the corresponding current is parallel to the stretched horizon.

Now we search for the other relations come from Maxwell equations and Bianchi identities. For the Maxwell equation we have,
\begin{equation}
\label{eq:14}
\partial_{\mu}(\frac{1}{g_{eff}^{2}}\sqrt{-g}F^{\mu\nu})=0,
\end{equation}
For simplicity,  we take effective coupling  $g_{eff}$ as a constant, so the components of the Maxwell equations will be,

\begin{equation}
\label{eq:15}
\sqrt{-g}g^{x_{i}x_{i}}g^{tt}\partial_{x_{i}}F_{x_{i}t}+\partial_{r}(\sqrt{-g}g^{rr}g^{tt}F_{rt})+\partial_{\theta}(\sqrt{-g}g^{\theta\theta}g^{tt}F_{\theta t})+\partial_{\Psi_{i}}(\sqrt{-g})g^{\Psi_{i}\Psi_{i}}g^{tt}F_{\Psi_{i}t}=0,
\end{equation}
\begin{equation}
\label{eq:16}
\sqrt{-g}g^{tt}g^{x_{i}x_{i}}\partial_{t}F_{tx_{i}}+\partial_{r}(\sqrt{-g}g^{rr}g^{x_{i}x_{i}}F_{rx_{i}})+\partial_{\theta}(\sqrt{-g}g^{\theta\theta}g^{x_{i}x_{i}}F_{\theta x_{i}})+\partial_{\Psi_{i}}(\sqrt{-g})g^{\Psi_{i}\Psi_{i}}g^{x_{i}x_{i}}F_{\Psi_{i}x_{i}}=0,
\end{equation}
\begin{equation}
\label{eq:17}
\sqrt{-g}g^{tt}g^{rr}\partial_{t}F_{tr}+\sqrt{-g}g^{rr}g^{x_{i}x_{i}}\partial_{x_{i}}F_{x_{i}r}+\partial_{\theta}(\sqrt{-g}g^{\theta\theta}g^{rr}F_{\theta r})+\partial_{\Psi_{i}}(\sqrt{-g})g^{\Psi_{i}\Psi_{i}}g^{rr}F_{\Psi_{i}r}=0,
\end{equation}
\begin{equation}
\label{eq:18}
\sqrt{-g}g^{tt}g^{\theta\theta}\partial_{t}F_{t\theta}+\sqrt{-g}g^{x_{i}x_{i}}g^{\theta\theta}\partial_{x_{i}}F_{x_{i}\theta}+\partial_{r}(\sqrt{-g}g^{\theta\theta}g^{rr}F_{r\theta })+\partial_{\Psi_{i}}(\sqrt{-g})g^{\Psi_{i}\Psi_{i}}g^{\theta\theta}F_{\Psi_{i}\theta}=0,
\end{equation}
\begin{equation}
\begin{split}
\label{eq:19}
&\sqrt{-g}g^{tt}g^{\phi\phi}\partial_{t}F_{t\phi}+\sqrt{-g}g^{x_{i}x_{i}}g^{\phi\phi}\partial_{x_{i}}F_{x_{i}\phi}+\partial_{r}(\sqrt{-g}g^{\phi\phi}g^{rr}F_{r\phi })  \\
 &+\partial_{\theta}(\sqrt{-g}g^{\phi\phi}g^{\theta\theta}F_{\theta\phi })+\partial_{\Psi_{i}}(\sqrt{-g})g^{\Psi_{i}\Psi_{i}}g^{\phi\phi}F_{\Psi_{i}\phi}=0,
\end{split}
\end{equation}

\begin{equation}
\begin{split}
\label{eq:20}
&\sqrt{-g}g^{tt}g^{\Psi_{i}\Psi_{i}}\partial_{t}F_{t\Psi_{i}}+\sqrt{-g}g^{x_{i}x_{i}}g^{\Psi_{i}\Psi_{i}}\partial_{x_{i}}F_{x_{i}\Psi_{i}}+\partial_{r}(\sqrt{-g}g^{\Psi_{i}\Psi_{i}}g^{rr}F_{r\Psi_{i} }) \\
&+\partial_{\theta}(\sqrt{-g}g^{\theta\theta}g^{\Psi_{i}\Psi_{i}}F_{\theta \Psi_{i}}) +\partial_{\Psi_{j}}(\sqrt{-g})g^{\Psi_{j}\Psi_{j}}g^{\Psi_{i}\Psi_{i}}F_{\Psi_{j}\Psi_{i}}=0.
\end{split}
\end{equation}
In second step we use Bianchi identities to extract other relations between components of the field strength,
\begin{equation}
\label{eq:21}
F_{[\mu\nu,\lambda]}=0.
\end{equation}

We take the following ansatz for the gauge field,
\begin{equation}
\label{eq:22}
A_{\mu}=A_{\mu}(r)e^{-i\omega t+i\overrightarrow{q}.\overrightarrow{x}},
\end{equation}
which $\mu$ runs over $t,x,y,z,r,\theta,\Phi,\psi_{i}$. Without lose of generality we can fix $\overrightarrow{x}$ along the spatial direction $x$, so that,
\begin{equation}
\label{eq:23}
A_{\mu}=A_{\mu}(r)e^{-i\omega t+iqx}.
\end{equation}
So one of the Bianchi identities is,
\begin{equation}
\label{eq:24}
\partial_{r}F_{tx}+\partial_{x}F_{rt}+\partial_{t}F_{xr}=0,
\end{equation}
which in complementary to Maxwell equations helps us to find the diffusion constant.
By choosing $A_{r}= A_{\Psi_{i}}=0$  and applying $\mid\partial_{t}A_{x}\mid\ll\mid\partial_{x}A_{t}\mid$ (which will prove later) we extract other components of the vector gauge field,
\begin{equation}
\label{eq:25}
-q^{2}\sqrt{-g}g^{tt}g^{xx}A_{t}+\partial_{r}(\sqrt{-g}g^{rr}g^{tt}\partial_{r}A_{t})
+i\omega\partial_{\theta}(\sqrt{-g}g^{\theta\theta}g^{tt})A_{\theta}=0.
\end{equation}

\begin{equation}
\label{eq:26}
\begin{split}
-\omega q\sqrt{-g}g^{tt}g^{xx}A_{t}+\partial_{r}(\sqrt{-g}g^{rr}g^{xx}\partial_{r}A_{x})
-iq\partial_{\theta}(\sqrt{-g}g^{\theta\theta}g^{xx})A_{\theta}=0,
\end{split}
\end{equation}

\begin{equation}
\label{eq:27}
i\omega\sqrt{-g}g^{rr}g^{tt}\partial_{r}A_{t}-iq\sqrt{-g}g^{rr}g^{xx}\partial_{r}A_{x}
-\partial_{\theta}(\sqrt{-g}g^{\theta\theta}g^{rr})\partial_{r}A_{\theta}=0.
\end{equation}

The differential equations for other components of Maxwell equations can be decoupled and the result is,
\begin{equation}
\partial_{r}(\sqrt{-g}g^{rr}g^{yy}\partial_{r}A_{y})-\omega^{2}\sqrt{-g}g^{tt}g^{yy}A_{y}=0
\end{equation}

\begin{equation}
\partial_{r}(\sqrt{-g}g^{rr}g^{zz}\partial_{r}A_{z})-\omega^{2}\sqrt{-g}g^{tt}g^{zz}A_{z}=0
\end{equation}

\begin{equation}
\label{eq:14}
\partial_{r}(\sqrt{-g}g^{\theta\theta}g^{rr}\partial_{r}A_{\theta})-(\omega^{2}\sqrt{-g}g^{tt}g^{\theta\theta}+q^{2}\sqrt{-g}g^{xx}g^{\theta\theta})A_{\theta}=0,
\end{equation}

\begin{equation}
\label{eq:28}
\partial_{r}(\sqrt{-g}g^{\phi\phi}g^{rr}\partial_{r}A_{\phi})-(\omega^{2}\sqrt{-g}g^{tt}g^{\phi\phi}+q^{2}\sqrt{-g}g^{xx}g^{\phi\phi})A_{\phi}=0,
\end{equation}

Now we try  to solve the above independent second order differential equations up to $O(\frac{l^{2}}{r_{0}^{2}})$ at stretched horizon ($r_{h}$) and in the equatorial plan i.e. $\theta=\frac{\pi}{2}$.
For simplicity, we apply the variable change $\frac{r_{h}-r_{e}}{r_{e}}=\varepsilon \ll 1,$ which helps us to expand the coefficients of the differential equation in powers of $\varepsilon$.

From Eq.(28) one can find $A_{y}$ as,
\begin{equation}
\label{29}
\varepsilon^2\partial_{\varepsilon}^{2}A_{y}+M\varepsilon\partial_{\varepsilon}A_{y} +\Omega^{2}A_{y}=0,
\end{equation}
where $M\equiv1+\frac{37}{16}\frac{l^2}{r_{0}^{2}}$ and $\Omega^{2}\equiv\frac{\omega^{2}cosh^{2}\alpha r_{0}^2}{16}(1+\frac{l^{2}}{2r_{0}^{2}})$.  This equation is the Euler equation and it's physical solution would be,
\begin{equation}
\label{eq:30}
A_{y}(\varepsilon)=C_{1}\varepsilon^{-\Lambda+ i\Omega}=C_{1}\varepsilon^{-\Lambda}e^{i\Omega ln\varepsilon},
\end{equation}
where $\Lambda\equiv\frac{37}{32}\frac{l^2}{r_{0}^{2}}$ and $C_{1}$ is a constant which can be determined after normalization.
The differential equation for the component $A_{z}(\varepsilon)$ is exactly the same as that for $A_{y}(\varepsilon)$. So it will be,
\begin{equation}
\label{eq:31}
A_{z}(\varepsilon)=C_{2}\varepsilon^{-\Lambda+ i\Omega}=C_{2}\varepsilon^{-\Lambda}e^{i\Omega ln\varepsilon}.
\end{equation}
where $C_{2}$ is normalization constant.
On the other hand we try to determine the component $A_{\theta}$. Under the same conditions and constraints supposed for the case of $A_{y}(\varepsilon)$ and $A_{z}(\varepsilon)$ we can find,
\begin{equation}
\label{eq:32}
A_{\theta}(\varepsilon)=C_{3}\varepsilon^{-\Lambda+i\Omega}=C_{3}\varepsilon^{-\Lambda}e^{+i\Omega ln\varepsilon}.
\end{equation}
where $C_{3}$ is the normalization constant again. After some calculations the differential equation for the $A_{\phi}$ is,
\begin{equation}
\label{33}
\varepsilon^2\partial_{\varepsilon}^{2}A_{\phi}-W\varepsilon\partial_{\varepsilon}A_{\phi} -\Omega^{2}A_{\phi}=0,
\end{equation}
where $W=\frac{43}{16}\frac{l^{2}}{r_{0}^{2}}$ and it's solution up to second order perturbation is,
\begin{equation}
\label{eq:34}
A_{\Phi}(\varepsilon)=C_{4}\varepsilon^{\widehat{\Lambda}+i\widehat{\Omega}}=C_{4}\varepsilon^{\widehat{\Lambda}}e^{+i\widehat{\Omega} ln\varepsilon}.
\end{equation}
where $\widehat{\Lambda}\equiv\frac{1}{2}+\frac{43}{32}\frac{l^{2}}{r_{0}^{2}}$ and $\widehat{\Omega}\equiv\frac{1}{2}(\frac{\omega^{2}cosh^{2}\alpha r_{0}^{2}}{4}-1)^{\frac{1}{2}}(1+\frac{1}{2}\frac{43-\omega^{2}cosh^{2}\alpha r_{0}^{2}}{4-\omega^{2}cosh^{2}\alpha r_{0}^{2}}\frac{l^{2}}{r_{0}^{2}})$.

The $A_{t}$ component of the gauge field could be determined from Eq.(25) for $\theta=\frac{\pi}{2}$ in which the third term vanishes. So, we have,
\begin{equation}
\label{eq:35}
-q^{2}\sqrt{-g}g^{tt}g^{xx}A_{t}+\partial_{r}(\sqrt{-g}g^{rr}g^{tt}\partial_{r}A_{t})=0.
\end{equation}
this equation can be written as,
 \begin{equation}
\label{eq:36}
\partial_{r}^{2}A_{t}+\partial_{r}ln(\sqrt{-g}g^{rr}g^{tt})\partial_{r}A_{t}-q^{2}\frac{g^{xx}}{g^{rr}}A_{t}=0.
\end{equation}
To find a solution for this equation, we first take $q=0$, so that the solution will be,
\begin{equation}
\label{eq:37}
A_{t}(r)=C_{0}\int_{r}^{\infty}\frac{dr'}{\sqrt{-g(r')}g^{rr}(r')g^{tt}(r')}
\end{equation}
In the second step, we assume $q$ is nonzero and small. After some calculations, we have $q^{2}\frac{g^{xx}}{g^{rr}}\sim q^{2}(\frac{1}{T^{2}}+l^2)$. This shows that (\ref{eq:36}) can be solved perturbatively when we take $\frac{q^2}{T^2}\ll1$ and $q^{2}l^2\ll1$. So $A_{t}$ can be written as a series on $\frac{q^{2}}{T^{2}}$,
\begin{equation}
\label{eq:38}
A_{t}=A_{t}^{(0)}+A_{t}^{(1)}+... \;\;\;\;\;\;\;,\;\;\;\;\;\; A_{t}^{(1)}=O(\frac{q^{2}}{T^{2}}),
\end{equation}
so we can write $A_{t}$ such that it has the same r-dependance and $C_{0}$ depends on $t$ and $x$. So,
 we have,
\begin{equation}
\label{eq:39}
A_{t}(r)=C_{0}(t)e^{iqx}\int_{r}^{\infty}\frac{dr'}{\sqrt{-g(r')}g^{rr}(r')g^{tt}(r')}.
\end{equation}
To find the diffusion constant, we need $\frac{A_{t}}{F_{tr}}\mid_{r_{e}}$ which after a lengthy but strightforward calculation will be as  follows,
\begin{equation}
\label{eq:40}
\frac{A_{t}}{F_{tr}}\mid_{r_{e}}=\frac{-r_{0}cosh^{2}\alpha}{2cosh^{2}\alpha+4}\left(1+\frac{l^{2}}{r_{0}^{2}}\frac{cosh^{2}\alpha}{9cosh^{2}\alpha+4}
(\frac{57}{4}+\frac{3}{2}cosh^{2}\alpha+\frac{3}{cosh^{2}\alpha}-\frac{4}{cosh^{4}\alpha})\right).
\end{equation}
This ratio will be used at the next section.

The last component of the vector gauge field is $A_{x}$ which can be determined explicitly from Eq.(27). The third term in that equation vanishes as we are on $\theta=\frac{\pi}{2}$, so we have,
\begin{equation}
\label{eq:41}
-\omega g^{tt}\partial _{r}A_{t}+qg^{xx}\partial_{r}A_{x}=0.
\end{equation}
By considering (42) and $A_{x}\mid_{r=\infty}=0$ we have,
\begin{equation}
\label{eq:42}
A_{x}(r)=\frac{\omega}{q}C_{0}(t)e^{iqx}\int_{r}^{\infty}\frac{dr'}{\sqrt{-g(r')}g^{rr}(r')g^{xx}(r')}.
\end{equation}
After some calculations it proves that for very small $r\rightarrow r_{0}$ we have,
\begin{equation}
\label{eq:43}
A_{x}\sim A_{t}\frac{\omega}{q}\sqrt{\frac{r_{0}}{r-r_{0}}},
\end{equation}
Therefor we can write,
\begin{equation}
\label{eq:44}
\frac{\mid\partial_{t}A_{x}\mid}{\mid\partial_{x}A_{t}\mid}\sim \frac{\omega^{2}}{q^{2}}\sqrt{\frac{r_{0}}{r-r_{0}}}.
\end{equation}
Since $\frac{\omega^{2}}{q^2}\sim\frac{q^{2}}{T^2}\ll1$, to let the above fraction to be much smaller than 1 we have,
\begin{equation}
\label{eq:45}
\sqrt{\frac{r_{0}}{r-r_{0}}}\ll \frac{T^{2}}{q^2},
\end{equation}
which means $r-r_{0}$ can not be too small and,
\begin{equation}
\label{eq:49}
\frac{r-r_{0}}{r_{0}}\gg \frac{q^{4}}{T^4}.
\end{equation}
Note that to put the stretched horizon on this region have taken in (\ref{eq:44}) the constraint $|\partial_{t}A_{x}|\ll\mid\partial_{x}A_{t}\mid$. Also we will find an upper bound for the location of the stretched horizon in the next section.
After now, we try to calculate the diffusion constant of slowly rotating three-brane by establishing a suitable connection between components of the Maxwell equations and Bianchi identities.

\section{Fick's first law and diffusion constant  }
In this section, we try to extract Fick's first law by applying previous results and conditions. First of all we consider Eq.(17) which have four terms. The fourth term vanishes because of the gauge fixing ($A_{r}=A_{\Psi_{i}}=0$). We restrict solutions to be in the equatorial plane, i.e. $\theta=\frac{\pi}{2}$. This, in turn, will eliminate the third term. Thus we have,
\begin{equation}
\label{eq:50}
\sqrt{-g}g^{tt}g^{rr}\partial_{t}F_{tr}+\sqrt{-g}g^{rr}g^{xx}\partial_{x}F_{xr}=0.
\end{equation}
By taking derivative with respect to $t$ then  combining it with Eq.(24) we have,
\begin{equation}
\label{eq:51}
\partial_{t}^2 F_{tr}+\frac{g^{xx}}{g^{tt}}(-iq\partial_{r}F_{tx}+q^{2}F_{rt})=0.
\end{equation}
 which by some calculations we arrive at,
\begin{equation}
\label{eq:52}
\partial_{t}^{2}F_{tr}-4\varepsilon\left\{1-(1-\frac{1}{2cosh^{2}\alpha})\frac{l^{2}}{r_{0}^{2}}\right\}\left\{-iq\partial_{r}F_{tx}+q^{2}F_{rt}\right\}=0,
\end{equation}
which is equivalent to,
\begin{equation}
\label{eq:53}
F_{tr}+4\varepsilon\left\{1-(1-\frac{1}{2cosh^{2}\alpha})\frac{l^{2}}{r_{0}^{2}}\right\}\left\{\frac{-iq}{\omega^{2}}\partial_{r}F_{tx}+\frac{q^{2}}{\omega^{2}}F_{rt}\right\}=0.
\end{equation}
Note that $\varepsilon\frac{q^2}{\omega^{2}}=\frac{r-r_{0}}{r_{0}}\frac{q^2}{T^2}$ and therefor we choose,
\begin{equation}
\label{eq:54}
\frac{r-r_{0}}{r_{0}}\frac{q^2}{\omega^2}\ll 1, \;\;\;\;\;\;\;\rightarrow \;\;\;\;\;\;\;\;\;\frac{r-r_{0}}{r_{0}}\ll \frac{\omega^{2}}{q^{2}}\sim\frac{q^2}{T^2}.
\end{equation}
This relation gives an upper bound on the location of the stretched horizon and can be satisfied simultaneously with Eq. (49). By these constraints we can write,
\begin{equation}
\label{eq:55}
F_{tr}\sim4\varepsilon \left\{1-(1-\frac{1}{2cosh^{2}\alpha})\frac{l^{2}}{r_{0}^{2}}\right\}\frac{q}{\omega^{2}}\partial_{r}F_{tx}.
\end{equation}
On the other hand, by taking derivative of Eq.(16) with respect to $t$, we arrive at
\begin{equation}
\label{eq:56}
\partial_{t}^{2}F_{tx}+\frac{1}{\sqrt{-g}g^{tt}g^{xx}}\partial_{r}\left\{\sqrt{-g}g^{rr}g^{xx}(\partial_{r}F_{tx}+\partial_{x}F_{rt})\right\}=0.
\end{equation}
Now by inserting Eq.(55) into Eq.(56) and taking the long wavelength regime into account, it is recognized that the third term of the Eq.(56) can be neglected. So, we fined the governing wave equation for $F_{tx}$ as follows,
\begin{equation}
\label{eq:57}
\partial_{t}^{2}F_{tx}+\frac{1}{\sqrt{-g}g^{tt}g^{xx}}\partial_{r}\left\{\sqrt{-g}g^{rr}g^{xx}\partial_{r}F_{tx}\right\}=0.
\end{equation}
Straightforward calculations will result in the following solution for the differential equation (57),
\begin{equation}
\label{eq:58}
F_{tx}(t,\varepsilon)=C_{7}e^{-i\omega t}\varepsilon^{-\Lambda+ i \Omega}=C_{7}e^{-i\omega t}\varepsilon^{-\Lambda}e^{ i \Omega ln\varepsilon},
\end{equation}
from which we have,
\begin{equation}
\label{eq:59}
\partial_{r}F_{tx}=\frac{i(-\Lambda+ i \Omega)}{\omega r_{e}\varepsilon}\partial_{t}F_{tx}.
\end{equation}
By inserting this relation into Bianchi identities we obtain,
\begin{equation}
\label{eq:60}
\partial_{t}\left[F_{rx}-\frac{i(-\Lambda+ i \Omega)}{\omega r_{e}\varepsilon}F_{tx}\right]+\partial_{x}F_{tr}=0.
\end{equation}
This equation has a solution only when the terms in the parenthesis be independent of $t$, and further be zero, in order to give a finite solution  at $t\rightarrow\infty$. So we have,
\begin{equation}
\label{eq:61}
F_{rx}=\frac{i(-\Lambda+ i \Omega)}{\omega r_{e}\varepsilon}F_{tx},
\end{equation}

Now we turn back to the currents. As it was calculated in previous section, the current along the spatial direction is,
\begin{equation}
\label{eq:62}
J^{x}=\frac{1}{g_{xx}\sqrt{g_{rr}}}F_{xr},
\end{equation}
By applying Eq.(61) we find,
\begin{equation}
\label{eq:63}
J^{x}=\frac{-1}{g_{xx}\sqrt{g_{rr}}}\frac{i(-\Lambda +i \Omega)}{\omega r_{e}\varepsilon}F_{tx},
\end{equation}
Using our previous condition $|\partial_{t}A_{x}|\ll\mid\partial_{x}A_{t}\mid$, the equation (63) will be,
 \begin{equation}
\label{eq:64}
J^{x}=\frac{1}{g_{xx}\sqrt{g_{rr}}}\frac{i(-\Lambda+ i \Omega)}{\omega r_{e}\varepsilon}\left(\frac{A_{t}}{F_{tr}}\right) \partial_{x}F_{tr},
\end{equation}
which can be recast into,
\begin{equation}
\label{eq:65}
J^{x}=\frac{-1}{g_{xx}{g^{tt}}}\frac{(-i\Lambda- \Omega)}{\omega r_{e}\varepsilon}\left(\frac{A_{t}}{\partial_{r}A_{t}}\right)\partial_{x}J^{t}.
\end{equation}
So we have finally,
\begin{equation}
\label{eq:66}
\begin{split}
J^{x}=-D\partial_{x}J^{t},
\end{split}
\end{equation}
which is the first Fick's law and $D$ is the diffusion constant. So the diffusion constant of the rotating three-brane will be,
\begin{equation}
\label{eq:67}
\begin{split}
D=\textit{D}_{(static)}+\frac{l^{2}}{r_{0}^{2}}\left({\textit{D}_{r_{(rotating)}}}+i{\textit{D}_{i_{(rotating)}}}\right),
\end{split}
\end{equation}
in which $\textit{D}_{(static)}$ is the static part of the diffusion constant while ${\textit{D}_{r_{(rotating)}}}$ and ${\textit{D}_{i_{(rotating)}}}$  are real and imaginary parts of diffusion constant respectively. ${\textit{D}_{r_{(rotating)}}}$ and ${\textit{D}_{i_{(rotating)}}}$ vanishes for the case of static spacetimes.
\begin{equation}
\label{eq:68}
\begin{split}
&\textit{D}_{(static)}=\frac{r_{0}cosh^{3}\alpha}{2cosh^{2}\alpha+4}, \\
&\textit{D}_{r_{(rotating)}}=\frac{r_{0}cosh^{3}\alpha}{2cosh^{2}\alpha+4}\left(\frac{\frac{3}{2}cosh^{6}\alpha+\frac{57}{4}cosh^{4}\alpha+
3cosh^{2}\alpha-4}{9cosh^{4}\alpha+4cosh^{2}\alpha}+\frac{1}{2cosh^{2}\alpha}-\frac{1}{2}\right), \\
&\textit{D}_{i_{(rotating)}}=\frac{r_{0}cosh^{3}\alpha}{2cosh^{2}\alpha+4}(\frac{37}{8\omega cosh\alpha r_{0}}),
\end{split}
\end{equation}
if we take $cosh\alpha\gg1$, i.e. at large $N$ limit, we have,
\begin{equation}
\label{eq:69}
 D=\frac{r_{0}cosh\alpha}{2}\left\{1+\frac{l^{2}}{r_{0}^{2}}\frac{cosh^{2}\alpha}{6}\right\},
\end{equation}
in this limit, the imaginary part of the diffusion constant can be neglected compared with the real part. So we can write diffusion constant in terms of temperature as follows,
\begin{equation}
\label{eq:70}
D=\frac{1}{2\pi T}\left\{1+\frac{l^{2}}{r_{0}^{2}}\frac{cosh^{2}\alpha}{6}\right\}.
\end{equation}
It is obvious that when rotation goes to zero, Eq.(70) agrees with results from the literature exactly ~\cite{o}.

\section{Fick's second law and quasinormal modes}
In the framework of general relativity, perturbations of black hole spacetimes can produce quasinormal modes (QNM). Particularly the gravitational waves which emit from black holes have quasinormal frequencies with real and imaginary parts~\cite{ll,lll,llll}. For higher dimensional black holes, the spectra of gravitational QNMs determines the stability of black branes, particularly when all QNMs are damped the stability is guaranteed \cite{Konoplya}.

 By considering gauge/gravity duality these modes set the stage for description of the hydrodynamic regime in the dual finite temperature strongly coupled quantum field theory. It proves that the lowest quasinormal frequencies of black branes can be interpreted as dispersion relations of hydrodynamic excitations in the dual theory.

 Perturbed Einstein equation is a set of coupled equations for higher dimensional black holes . To solve the equations three channels of perturbations can be considered: scalar, shear and sound channel. Each type of these channels perturbs particular components of metric. Then three independent second order differential equation will obtain. The hydrodynamic modes dependency on $q$ extracts from the solutions of the corresponding differential equations.

 At the field theory side, the poles of the retarded two-point function of the stress energy tensor give information about speed of sound and shear and bulk viscosity. Dispersion relations for the shear and sound modes emerges from small deviation of $T_{\mu\nu}$ from equilibrium. The conservation law $ \partial_{\mu}T^{\mu\nu}=0$ together with the constitutive relations yields equations of linearized hydrodynamics. Some calculations for the low frequency, small momenta fluctuations of the stress energy tensor of any $d$-dimensional theory (see (139) in \cite{ll}) gives two type of eigen modes, the shear mode with the dispersion relation,
\begin{equation}
\label{eq:71}
\omega=-i\frac{\eta}{p+\epsilon}{q}^2+O(q^4),
\end{equation}
and the sound mode with the dispersion relation,
\begin{equation}
\label{eq:72}
\omega=\pm C_s q -i\frac{1}{p+\epsilon}\left[\frac{\xi}{2}+(1-\frac{1}{d})\eta\right]q^2+O(q^3).
\end{equation}
where $\epsilon$, $p$, $\eta$ ,  $\xi$ are energy density, pressure, shear and bulk viscosity respectively and $C_s$ is the speed of sound.
This equation leads us to an overdamped mode which emerges from the real part of the diffusion constant and shows the diffusion of the conserved charge.

At the other hand, the Fick's first law, $J^{x}=-D\partial_{x}J^{t}$, with conservation equation of current $\partial_{\mu}J^{\mu}=0$ yields Fisk's second law, $\partial_{t}J^{t}=D\partial_{x}^{2}J^{t}$, which is,
  \begin{equation}
  \label{eq:71}
  \omega=-iDq^2,
  \end{equation}
 so the linearized hydrodynamics predicts the existence of a simple pole with above dispersion relation.

The relation (73) exists for all static or stationary black holes. But extracting QNMs for the case of rotating black holes is now under consideration in the literature. Analytical solutions, even perturbative methods, are very challenging and have not been done yet. Some recent progress has been made in the subject of the scalar field perturbations of the rotating black holes \cite{Yoshida,Hod}. For the case of Maxwell field perturbations, finding analytic solutions are still under exploration (For some restricted numerical computations see \cite{Wang1,Wang2}). These last papers have found quasi-frequencies which have both real and imaginary parts, in agreement to our result. In other words, having a damping factor emerging from the imaginary part of the mode frequencies is a typical behaviour of the rotating black holes (see the above references), a phenomenon we hope to appears also for black branes. In fact, using the obtained diffusion constant and the Fick's second law, $\omega=-iDq^2$, one finds $\omega$ explicitly. To confirm our result, a separate (and difficult) research must be devoted to find directly $\omega$, using traditional techniques like continued fractions method, motion groups technique for solving the differential equations and so on.

\section{Conclusion and outlook }

 In this paper we studied the spacetime of slowly rotating black three-brane in ten dimensional spacetime. By taking nontrivial supergravity background and using holographic theory we have obtained an explicit expression for the diffusion constant of the corresponding dual field theory. Specially we made long wavelength description of the dual field theory which we expected to be connected to hydrodynamics.

 Relating different components of Maxwell equations and Bianchi identities, we realized Fick's first law. Extracting Fick's first law leads us to determine diffusion constant explicitly. Surprisingly, the diffusion constant which we obtained for the rotating black three-brane had both real and imaginary parts. It is obvious that imaginary term has been appeared because of the rotation of the black brane. If we take static limit, the diffusion constant will be real, in agreement with the literature. Also, second law of Fick helped us to find dispersion relation for the dual theory. The real part of the diffusion shows dispersion of the conserved charge and damping of the hydrodynamic flow. The imaginary part corresponds to the oscillating propagation of the gauge field.

 In order to provide some supports for the presence of the complex constant $D$, we employed the relation between QNM and  diffusion constant. It will be interesting to explain, as a separate research, the stability and un-stability of rotating three-brane using the QNM spectra relation and Eq.(67).
 Some types of QNMs for rotating black holes which has been calculated in the literature are complex (see Refs. (27)-(30)). This type of QNMs supports indirectly our result, i.e. having a complex diffusion constant. The viscosity bound is sometimes violated, for example in the supergravity theory with higher derivative terms because of the corrections to the QNM spectra. Thus obtaining QNMs of rotating black three-brane and its probable corrections determines whether the viscosity bound saturates or may has corrections.        \\


\begin{thebibliography}{99}
%\bibliography{mybibfile}

\bibitem{a}
G. T. Horowitz and A. Strominger, \emph{Black strings and p-branes}, Nucl. Phys. B 360 (1991) 197.

\bibitem{b}
J. G. Russo, \emph{New compactifications of supergravities and large N QCD},  Nucl.Phys. B 543 (1999) 183 [hep-th/9808117].

\bibitem{c}
M. Cvetic and D. Youm, \emph{Rotating intersecting M-branes}, Nucl. Phys. B499 (1997) 253 [hep-th/9612229].


\bibitem{d}
S.S. Gubser, \emph{Thermodynamics of spinning D3-branes}, Nucl. Phys. B 551 (1999) 667 [hep-th/9810225].

\bibitem{e}
G. 't Hooft, \emph{Dimensional reduction in quantum gravity},  Salamfest (1993)0284 [gr-qc/9310026].

\bibitem{f}
L. Susskind, \emph{The world as a hologram}, J. Math. Phys. 36 (1995) 6377 [hep-th/9409089].

\bibitem{g}
R. Bousso, \emph{The holographic principle}, Rev. Mod. Phys. 74 (2002) 825 [hep-th/0203101].

\bibitem{h}
G.Policastro, D. T. Son and A. O. Starinets, \emph{Shear viscosity of strongly coupled N=4 supersymmetric Yang-Mills plasma}, Phys. Rev. Lett. 87 (2001) 081601 [hep-th/0104066].

\bibitem{i}
S. Bhattacharyya, Veronika E. Hubeny, S. Minwalla, M. Rangamani, \emph{Nonlinear Fluid Dynamics from gravity},  JHEP 0802:045, (2008), [hep-th:0712.2456].

\bibitem{j}
M. Rangamani, \emph{Gravity and hydrodynamics: Lectures on the fluid-gravity correspondence}, Class. Quant. Grav. 26 (2009) 224003, [hep-th/0905.4352].

\bibitem{k}
G. Policastro, D.T. Son and A.O. Starinets, \emph{From AdS/CFT correspondence to hydrodynamics},  JHEP. 09 (2002) 043 [hep-th/0205052].

\bibitem{l}
G. Policastro, D.T. Son and A.O. Starinets, \emph{From AdS/CFT correspondence to hydrodynamics, II: Sound waves},  JHEP. 12 (2002) 054 [hep-th/0210220].

\bibitem{m}
C. P. Herzog, \emph{The hydrodynamics of M-theory}, JHEP. 12 (2002) 036 [hep-th/0210126].

\bibitem{n}
C. P. Herzog, \emph{The sound of M-theory}, Phys. Rev. D 68 (2003) 024013 [hep-th/0302086].

\bibitem{rr}
K. S Thorne, R. H. Price and D. A. Macdonald, \emph{Black holes: the membrane paradigm}, Yale University Press, New Haven 1986

\bibitem{ss}
M. Parikh and F. Wilczek, \emph{an action for black hole membranes}, phys. Rev. D 58 (1998) 064011 [gr-qc/9712077]


\bibitem{o}
P. Kovtun, Dam T. son and Anderi O. Starinets \emph{Holography and hydrodynamics: diffusion on stretched horizon}, JHEP 0310:064,2003 ,[hep-th/0309213].

\bibitem{gg}
T. Moskalets and A. Nurmagambetov, \emph{Liouville mode in Gauge/Gravity Duality},  Eur.Phys.J. C75 (2015) 11, 551, [hep-th/1409.4186]


\bibitem{cc}
J. C. Breckenridge, R. C. Myers, A. W. Peet and C. Vafa \emph{D-branes and spinning Black holes}, Phys.Lett. B391 (1997) 93-98, [hep-th/9602065]

\bibitem{ee}
M. Cvetic and Steven S. Gubser\emph{ Thermodynamic Stability and Phases of General Spinning Branes}, JHEP 9907:010,1999, [hep-th/9903132]

\bibitem{ff}
T. Harmark and N.A. Obers, \emph{Thermodynamics of Spinning Branes and their Dual Field Theories}, JHEP0001:008,2000, [hep-th/9910036]


\bibitem{dd}
J. G. Russo and K. Sfetsos \emph{Rotating D3 branes and QCD in three dimensions}, [hep-th/9901056].

\bibitem{ll}
E. Berti, V. Cardoso and A. O. Starinets, \emph{Quasinormal modes of black holes and black brane}, Class. Quant. Grav. 26 (2009) 163001    [hep-th/0905.2975].

\bibitem{lll}
 R. Baier, P. Romatschke, D. T. Son, A. O. Starinets and M. A. Stephanov,\emph{Relativistic viscous hydrodynamics, conformal invariance, and holography}, JHEP 04, 100 (2008), [hep-th/0712.2451].

\bibitem{llll}
A. Parnachev and A. Starinets, \emph{The Silence of the Little Strings}, JHEP 10, 027 (2005), [hep-th/0506144].

\bibitem{Konoplya}
 R. A. Konoplya, A. Zhidenko, \emph{Quasinormal modes of black holes: from astrophysics to string theory}, Rev.Mod.Phys.83:793-836 (2011),[hep-th/1102.4014].

\bibitem{Yoshida}
S. Yoshida, N. Uchikata, and T. Futamase,  \emph{Quasinormal modes of Kerr–de Sitter black holes},
Phys. Rev. D 81, 044005 (2010).

\bibitem{Hod}
Shahar Hod, \emph{Spinning Kerr black holes with stationary massive scalar clouds: the large-coupling regime}, JHEP 01 (2017) 030, [hep-th/1612.02819].

\bibitem{Wang1}
M. Wang, C. Herdeiro, \emph{Maxwell perturbations on Kerr-anti-de Sitter: quasinormal modes, superradiant instabilities and vector clouds}, Phys. Rev. D 93, 064066 (2016), [gr-qc/1512.02262].

\bibitem{Wang2}
M. Wang, \emph{Boundary conditions for Maxwell fields in Kerr-AdS spacetimes}, Int.J.Mod.Phys. D25 (2016) no.09, 1641011.




\end{thebibliography}
\end{document}